%% file: ReviewCR.tex
\documentclass[preprint,12pt]{elsarticle}

\usepackage{amssymb}

\journal{Astroparticle Physics}

\begin{document}
\newcommand{\el}{\mbox{${\rm e^{-}}$ }}
\newcommand{\ps}{\mbox{${\rm e^{+}}$ }}

\begin{frontmatter}



\title{Multi messenger astronomy and CTA: TeV cosmic rays and
      electrons}


\author[1,2]{Piergiorgio Picozza}
\author[3]{Mirko Boezio}
\address[1]{INFN, Sezione di Rome ``Tor Vergata'', I-00133 Rome, Italy}
\address[2]{University of Rome ``Tor Vergata'', Department of
Physics,  I-00133 Rome, Italy} 
\address[3]{INFN, Sezione di Trieste, I-34149
Trieste, Italy}

\begin{abstract}
Cosmic rays are a sample of solar, galactic and extragalactic matter. 
Their origin and properties are one of the most
intriguing 
question in modern astrophysics. The
most energetic events and active objects in the Universe: supernovae
explosion, pulsars, relativistic jets, active galactic nuclei, have
been proposed as sources of cosmic rays 
although unambiguous evidences have still to be 
found. Electrons, while comprising $\sim 1\%$ of the cosmic radiation,
have unique features providing important information 
regarding the origin and
propagation of cosmic rays in the Galaxy that is not accessible from
the study of the cosmic-ray nuclear components due to their differing
energy-loss processes. 
In this paper we will analyse, discussing the experimental
uncertainties and challenges, the most recent 
measurements on cosmic-ray nuclei and, in
particular, electrons with energies from tens
of GeV into the 
TeV region. 
\end{abstract}


\end{frontmatter}



\include{Introduction}
\include{Composition}

\include{Elet}

\section{Acknowledgments}
We would like to thank the PAMELA Collaboration for providing some of the 
information included in this paper and E. Mocchiutti for helping in the
production of several figures.






\bibliographystyle{model1-num-names}
\bibliography{ReviewCR}







\end{document}

%% file: Introduction.tex
\section{Introduction}
\label{intro}

\begin{figure}[ht]
\includegraphics[width=25pc,height=0.5\textheight]{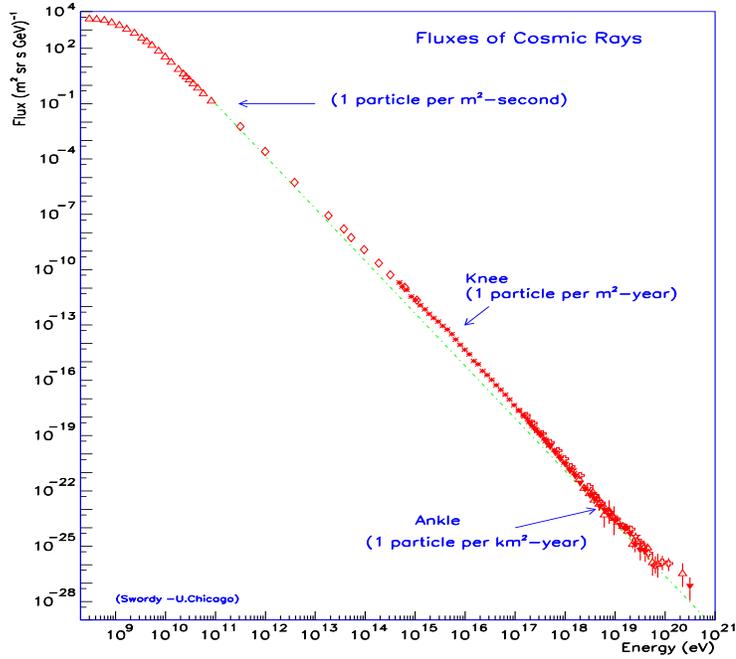}\hspace{2pc}
\caption{All particle cosmic ray energy spectrum. Figure
  from~\cite{cro97} 
\label{fig:spectra}}   
\end{figure}

Cosmic rays are a sample of solar, galactic and extragalactic matter
which includes all known nuclei and their isotopes, as well as
electrons, positrons, and antiprotons.  They are associated with the
most energetic events and active objects in the Universe: supernovae
explosion, pulsars, relativistic jets, active galactic nuclei,
although an unambiguous proof of their origin has not until now been
found. The measurement of all particle cosmic ray energy spectrum,
shown in Fig.~\ref{fig:spectra}, 
ranges for about 32 orders of magnitude in  flux
determination and more than 21 in  explored  energy.  Three features
appear in the spectrum, a first knee at an energy of 3 PeV, a second
knee at about 0.5 EeV and an ankle beyond 10 EeV. In this spectrum
there are, quite hidden, the answers to the main questions in the
cosmic rays research. Where do the particles are coming from? How and
where they are getting accelerated? How do they propagate through the
interstellar medium and what kind of interaction do they encounter?
What role do they play in the energy budget of the interstellar
medium? Are they galactic or also extragalactic?  The cosmic ray
particles, at least up to about 10$^{15}$ eV, are considered of galactic
origin and  shock waves of expanding supernovae remnants are ideal
candidates to supply  the power for their acceleration. 
Evidences supporting this hypothesis have recently been reported by
AGILE and  Fermi, with the 
observation of gammas presumably coming from neutral pion emission
from accelerated protons in the Supernova Remnant
W44~\cite{giu11,abd10} and SR IC 
443~\cite{tav10}, by Cherenkov imaging telescopes~\cite{fun08} and in
X-ray emissions 
from the borders of SRN \cite{vin05}. The accelerated particles  are injected
into the  interstellar space,  where they  remain about 10$^{7}$ years
before escaping into the intergalactic space. This long period is  due
to their coupling to the galactic magnetic  field  and  to scattering
on  random magnetic fields that leads to a random walk in real space
(diffusion) and  momentum space (diffusive reacceleration).  Diffusion
provides a high degree of isotropy for  the cosmic rays, however the
random nature (in space and time) especially of nearby recent SNRs
could produce the small anisotropy observed at Earth \cite{bla11b}.
Particles 
are  also spatially convected  away by the galactic wind,  inducing
adiabatic losses and lose energy when they interact with interstellar
matter or, especially for electrons,  with the electromagnetic field
and radiation of the Galaxy by synchrotron radiation  and inverse
Compton scattering.  New particles and  spallation products are
obtained by interaction of cosmic rays with the interstellar
matter. Solar modulation affects the low energy part of the cosmic
rays and plays an important role in the precise  determination of the
cosmic ray interstellar energy spectrum.  
Understanding the origin, acceleration  and  propagation of cosmic
rays in the Galaxy 
requires the combination of many different investigations over a wide
range of energy, including chemical composition, anisotropy, and solar
modulation. 
 

%% file: Composition.tex
\section{Cosmic rays elemental composition}
\label{Comp}

The experimental methods divide  the cosmic ray energy spectrum in two
large intervals. The first, below the first knee, is explored  by
direct measurements, carried out by experiments on stratospheric
balloons or in space on board satellites or International Space
Station, with  single particle identification and energy definition.
The second one, over the knee, where the low particle flux makes
feasible only indirect observations, is measured by very large on
ground detectors and the particle nature is inferred, on basis of  the
statistics and with considerable systematic uncertainties, studying
their interaction with the atmosphere. 
 The experimental observation of a change of slope, or ``knee'', in the
 cosmic ray flux at 10$^{15}$ eV energy scale has not yet found an
 explanation universally accepted, after almost half a century since
 its discovery.  According to some theoretical models,  the knee is
 linked to the process of acceleration of cosmic rays.  In case of
 acceleration due to shock waves of expanding supernovae remnants,
 the existence of a maximum rigidity R = pc/(Ze), p being the momentum
 of a particle of charge Ze, is predicted,  to which the mechanism of
 acceleration becomes inefficient.  In this scenario, the spectrum of
 each individual element in primary cosmic rays would show a "cutoff"
 at a characteristic momentum with a linear dependence  on the atomic
 number Z.  These models appear to be confirmed by experimental data
 that show a ``knee''  at $3 \div 5 \times 10^{15}$~eV presumably due to
 light primary 
 masses cutoff \cite{ant02,ant05}  and, very recently,  another at 
$8 \times 10^{16}$ eV
 attributed to heavy primary nuclei \cite{ape11}. Therefore, the observed
 inclusive spectrum  would be the superposition of individual spectra,
 weighted with the relative abundances of the elements in the flux of
 the primaries and the knee would reflect the different  composition
 of cosmic rays. An alternative explanation of the knee is adopted by
 models that relate it to leakage of cosmic rays from the Galaxy. In
 this case the knee is expected to occur at lower energies for light
 nuclei as compared to heavy ones, due to the rigidity-dependence of
 the Larmor radius of cosmic rays propagating in the galactic magnetic
 field \cite{hor04}. 

\begin{figure}[ht]
\begin{center}
\includegraphics[width=25pc,height=0.5\textheight]{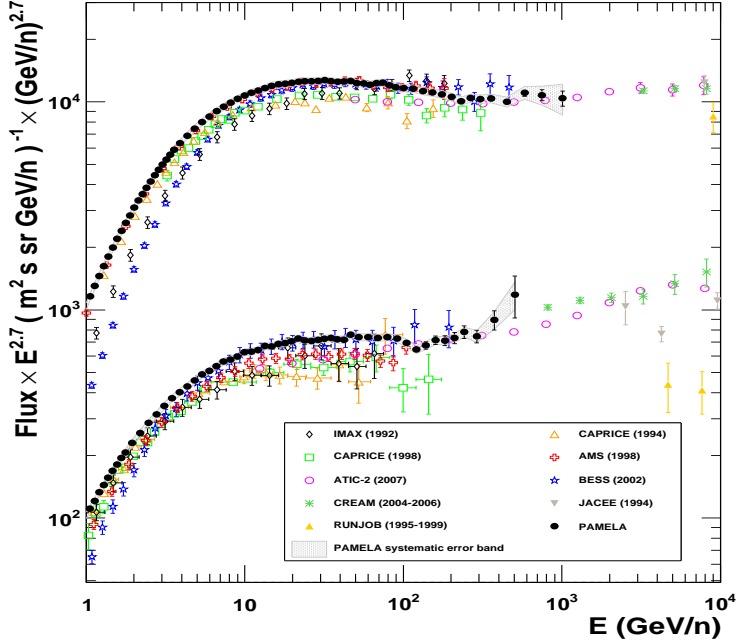}\hspace{2pc}
\end{center}
\caption{Proton and helium absolute fluxes  measured  between  1 GeV/n
  and 10 TeV/n: JACEE \cite{asa98}, CAPRICE 1994 \cite{boe99}, AMS-01
  \cite{alc00a}, IMAX \cite{men00}, CAPRICE
  1998 \cite{boe03}, BESS \cite{hai04}, RUNJOB
  \cite{har06}, ATIC-2
  \cite{wef08},  PAMELA \cite{adr11a}, CREAM \cite{yoo11}.   
The measurements have been obtained by balloon-borne experiments but
AMS-01 and PAMELA.} 
\label{fig:PHeflux}
\end{figure}

\begin{figure}[ht]
\begin{center}
\includegraphics[width=25pc,height=0.5\textheight]{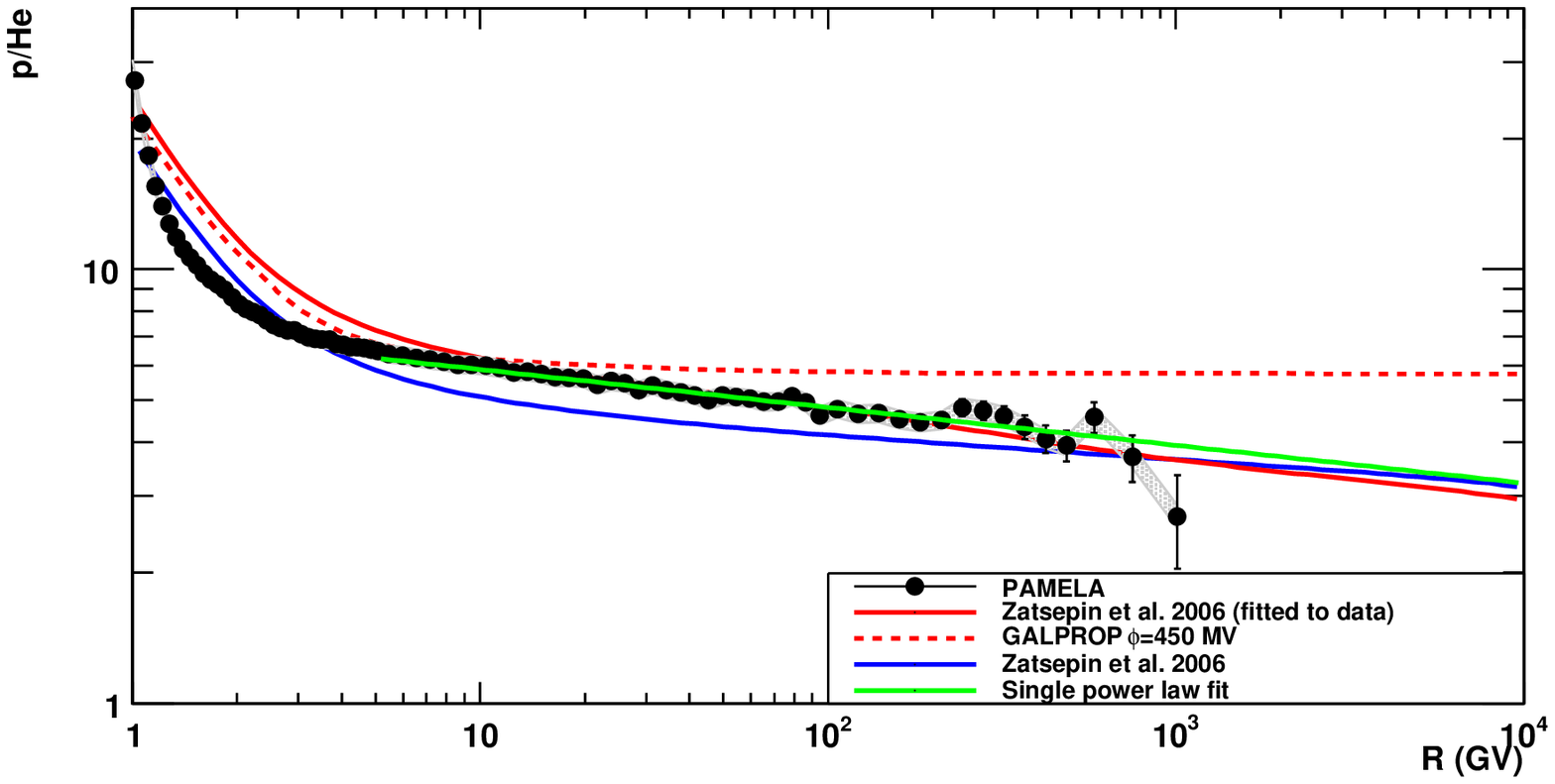}\hspace{2pc}
\end{center}
\caption{Ratio  between proton and helium data  of  PAMELA vs. Rigidity. The
shaded area represents the estimated systematic uncertainty. Lines
show the fit using one  single power law (describing the difference of
the two spectral indices), the GALPROP calculation and the Zatsepin
models using the original values of the paper~\cite{zat06} and fitted
to the 
data.} 
\label{fig:PHeflux2}
\end{figure}

\begin{figure}[ht]
\begin{center}
\includegraphics[width=25pc,height=0.5\textheight]{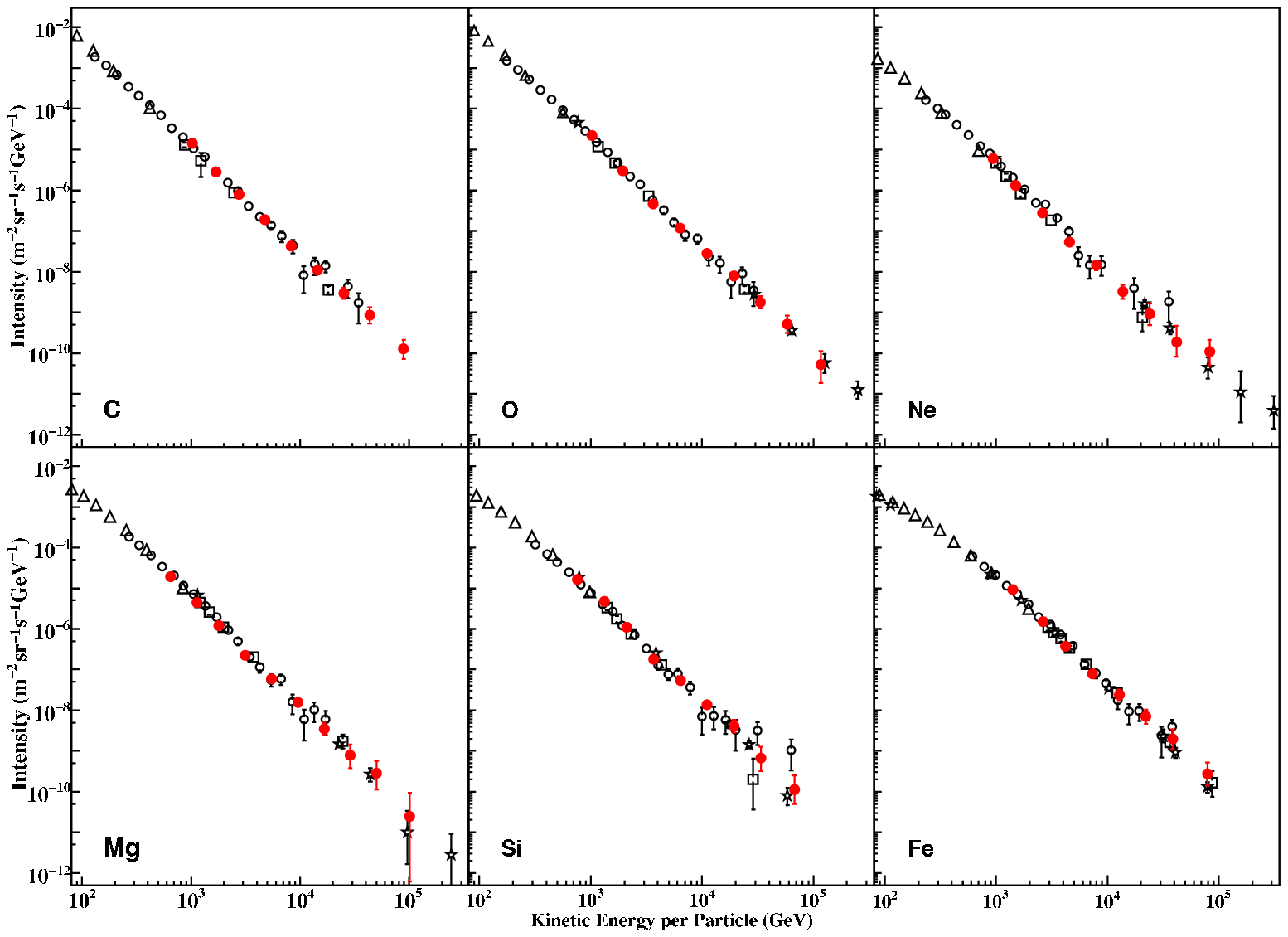}\hspace{2pc}
\end{center}
\caption{Energy spectra of the more abundant heavy nuclei. Results of
  CREAM-II \cite{mae09} (filled circles),  HEAO \cite{engelmann90}
  (triangles), CRN \cite{muller91} 
  (squares), ATIC \cite{pan07} (open circles) and TRACER \cite{ave08}
  (stars). Figure from~\cite{mae09}.}
\label{fig:Nuclei}
\end{figure}

In this paper we deal with direct measurements of cosmic rays under
the ``knee'', with a focus  on the chemical composition  that provides
important information about  cosmic ray  primary sources, secondary 
production, acceleration and propagation processes  through the
interstellar medium.  An extensive work has been conducted in space
and at the top of the atmosphere in the energy region between few
GeV/n and hundreds of TeV/n,  starting with the pioneer series of
Proton satellite experiments of Grigorov~\cite{gri71},  extended, later, by
the JACEE \cite{asa98} and RUNJOB \cite{har06} balloon-borne
experiments and  by HEAO \cite{engelmann90} and SOKOL \cite{iva93}
on board satellites and CRN \cite{muller91} on board the Space
Shuttle. At present 
new data are released by 
ATIC \cite{wef08}, BESS \cite{hai04}, CREAM \cite{yoo11}, TRACER
\cite{ave08}, PPB-BETS \cite{tor08} as 
results  of Long Duration  Balloon flights in Antarctica  and  by
PAMELA on satellite. Very precise measurement of nuclei  at
the TeV region are expected by AMS-02~\cite{bat05}, operating outside
the ISS since 
May 2011. 
The most precise data  on   proton  and  helium  energy spectra have
been achieved by PAMELA \cite{adr11a} between 1 GeV and 1.2 TeV  for protons
and 1 GeV/n  to 500 GeV/n  for helium and by CREAM \cite{yoo11}
between 2.5 
TeV and 250  TeV for protons and 630 GeV/n  to 63 TeV/n  for helium.
Results  from  different experiments
\cite{asa98,har06,wef08,hai04,yoo11,adr11a,boe99,men00,alc00a,boe03}
up to 10 TeV/ n, 
multiplied by E$^{2.7}$, are shown in Fig.~\ref{fig:PHeflux}. The
extrapolation to higher 
energy of the PAMELA fluxes suggests a good agreement with those
published by CREAM and JACEE~\cite{asa98} but they are higher than the RUNJOB
\cite{har06} helium data.   PAMELA data show a spectral hardening, at 230-240
GV, which appears present also in
the ATIC2 data \cite{wef08}.  
For long time a tantalizing question has focused on the possible
uniqueness of  the index spectrum for all nuclei, including protons.
It has been difficult to prove subtle differences between the
different spectra, because spectral indices determined by different
measurements performed over a limited energy range or with  low
statistics  and large background contamination could not provide a
definite answer. The PAMELA data, taken  over a wide energy range
above the atmosphere, show  clearly this difference between the proton
and helium slopes,  as seen in Fig.~\ref{fig:PHeflux2}, where the ratio
of the fluxes is 
presented as a function of rigidity. 
In this configuration the
possible impact of systematic errors is reduced, because several
instrumental effects cancel in the ratio. The proton-to-helium flux
ratio shows a continuous and smooth decrease and it is well described
by a power law down to rigidities as low as 5 GV  with a spectral
index of 0.1. The data are compared with Zatsepin model \cite{zat06}. 
The energy spectra of the most abundant heavy nuclei 
\cite{engelmann90,muller91,ave08,mae09} are
presented in Fig.~\ref{fig:Nuclei}. The agreement among the different
experiments 
appears to be quite good in the regions of the overlaps. The
insufficient precision of the measurements does not allow to observe a
significant trend of the spectral indices with Z charge.  Energy power
laws with an  average spectral index of  2.65 ± 0.05 fit well all the
TRACER data above 20 GeV \cite{ave08}. This behavior suggests a common
origin 
of all cosmic ray species. However, protons and helium data advise for
a more careful analysis. In fact, it is expected  that the competing
action of physical escape from the Galaxy, which depends on energy,
and  loss by spallation in the interstellar medium (ISM), which
depends on the nuclear charge Z (or more correctly, on atomic number
A), should lead to some changes in the spectral shape for the
different nuclei that  would be difficult to describe by the same
power-law spectrum. Moreover, a recent  accurate analysis~\cite{erl11} of
direct measurements of the energy spectra of the major mass components
of cosmic rays indicates  the presence of an 'ankle' in the region of
several hundred GeV/n. The ankle, which varies in magnitude from one
element to another, is much sharper than predicted by  cosmic ray
origin models in which supernova remnants are responsible for cosmic
ray acceleration and it appears as a new, steeper component is
necessary.

\subsection{Cosmic ray secondaries}

\begin{figure}[h]
\begin{center}
\includegraphics[width=25pc,height=0.5\textheight]{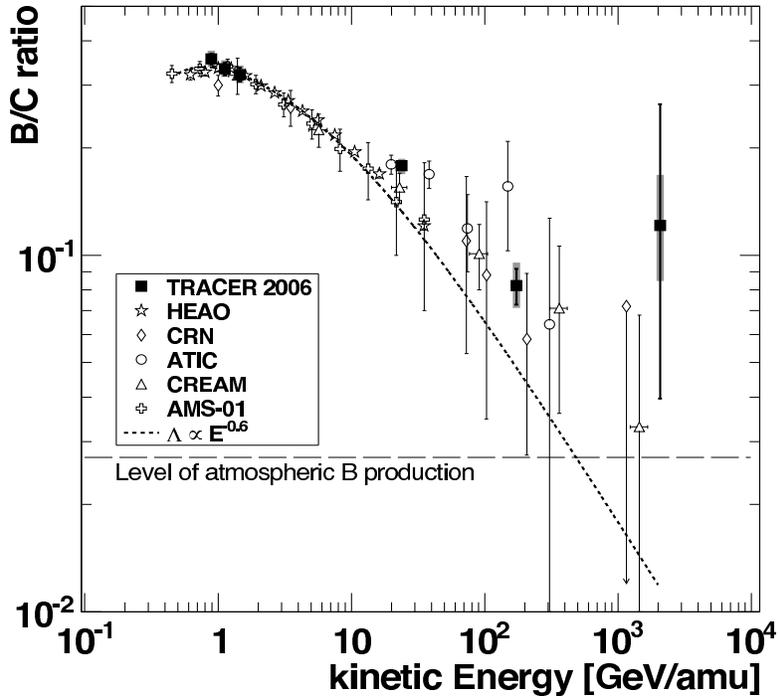}\hspace{2pc}
\end{center}
\caption{The boron-to-carbon abundance ratio as a function of kinetic
  energy per nucleon. 
Error bars are statistical (thin) and systematic (thick). A model
corresponding to an 
escape path length  E$^{-0.6}$ (dotted) is shown. The level of atmospheric
production of 
boron (dashed) is indicated.  HEAO \cite{engelmann90}, TRACER
\cite{obe11}, CRN  \cite{swordy90}, ATIC \cite{pan07}, CREAM
\cite{ahn08} and AMS-01 \cite{agu10}. Figure from~\cite{obe11}.} 
\label{fig:rationuclei}
\end{figure}

The combined effect of  acceleration and propagation of cosmic rays in
the Galaxy  leads  to a difference between the spectra  at  the source
and those measured at Earth.  Secondary nuclei are produced  by
spallation in the interaction of primary nuclei  with interstellar
matter.   Powerful tools  to characterize the diffusion property of
the ISM and to test the propagation models are  therefore the
measurements of  the abundances and energy spectra of secondary
elements such as Boron, Beryllium and Lithium and, particularly, the
ratios between secondary and primary cosmic ray fluxes as B/C, Be/C,
Li/C etc. They are  directly connected to the crossed amount of matter
in the Galaxy and to the nuclei lifetime before escaping from the
Galaxy. Actually, the energy dependence of the B/C ratio is directly
connected with the diffusion coefficient, or more in general with the
escape time, which scales as the inverse of the coefficient  if
diffusion is the only process responsible for escape. The results of
measurements of the B/C ratio performed  by several experiments 
\cite{engelmann90,pan07,obe11,swordy90,ahn08,agu10}
are shown in Fig.~\ref{fig:rationuclei}, that includes also an extrapolation
(dotted line) for an E$^{-0.6}$ decrease of this ratio with energy,
inferred from measurements at low energy.  However, this energy
dependence would imply  too small values for the escape path length of
cosmic rays at the highest energies and a large anisotropy of cosmic
rays at knee would be measured. Thus, it is quite interesting that the
more precise  data  of  TRACER \cite{obe11} on the B/C ratio above 100
GeV/nucleon, and also, as shown in Fig.~\ref{fig:rationuclei}, some of
the results of 
other measurements,  appear to lie above the E$^{-0.6}$ prediction. This
feature  may suggest that the energy dependence of  the escape  path
length flattens  at high energy, and perhaps indicates an asymptotic
transition to a constant residual path length. On the other hand, if
this transition exists  one should expect the appearance of a feature
in the all-particle spectrum  which does not seem to be there. 
In the next months, the AMS-02 space mission will allow for
fundamental improvements to the understanding of the origin and
propagation of cosmic rays in the Galaxy. New results at higher energy
will be provided by CALET experiment~\cite{tor11} planned to be
installed outside 
the ISS  in 2014.

%% file: Elet.tex
\section{High energy cosmic-ray electrons}
\label{electrondata}

Electrons (and positrons) constitute about 1\% of the total cosmic-ray
flux. While small, this component provides important information 
regarding the origin and
propagation of cosmic rays in the Galaxy that is not accessible from
the study of the cosmic-ray nuclear components due to their differing
energy-loss processes. In fact, because of their low mass, electrons
undergo severe energy losses through synchrotron
radiation in the Galactic magnetic field and inverse Compton scattering
with the interstellar radiation fields and the cosmic microwave
background radiation. 

There are two possible origins of high-energy electrons in the cosmic
radiation: primary electrons accelerated at sources such as SNR and
secondary electrons produced by processes such as nuclear interactions
of cosmic rays with the interstellar matter.
From an energetic
point of view it was realized already in the fifties
(e.g.~\cite{Gin64}) that supernovae
explosions  
released sufficient energy to power the cosmic rays in the
Galaxy.
Evidence for synchrotron X-ray emission 
\cite{koy95,all97} 
strongly supports the
hypothesis that 
Galactic cosmic-ray electrons 
originate in supernovae. In this case, TeV gamma rays should be
produced via the inverse Compton process between accelerated
electrons and the cosmic microwave background radiation and indeed TeV
gamma rays were detected (e.g.~\cite{aha04}).

Secondary electrons originate from the interaction of cosmic rays,
mostly protons and helium nuclei, on the interstellar matter (hydrogen
and helium). In this process positrons are produced too. Since the
interaction involves positively charged particles, charge conservation
implies that slightly more positrons than electrons are created
(e.g. \cite{kam06}). 
Since the observed positron component 
is of the order of ten percent and less of the electron one 
above a few GeV (e.g. \cite{des64}), the majority
of electrons must be of primary origin. 
However, additional
sources of electrons (and positrons) cannot be excluded. The recent
observation by the PAMELA satellite of a rising positron
fraction up to $\sim 100$~GeV \cite{adr09b}, later confirmed by the
Fermi experiment 
\cite{ack12} up to 200 GeV, has prompted a considerable effort in the 
theoretical interpretation of the data. New
sources ranging from astrophysical objects
such as pulsars, e.g.~\cite{mal09}, or 
more exotic sources such
as dark
matter particles, e.g.~\cite{cir08}, have been proposed to explain the
results. Similarly, the recent 
measurement of 
an excess in the ``all electron'' (\el + \ps) spectrum by the ATIC
collaboration~\cite{cha08}  at a few hundred GeV has been interpreted
in terms of a 
dark matter signal or a contribution of a nearby pulsar. 

Because of energy losses via synchrotron radiation and inverse
Compton scattering, the lifetime of high energy electrons is
approximately: $ 2.5 \times 10^{5} ({\rm yr}) / E ({\rm TeV})$
\cite{kob04}. In diffusion models for transport of cosmic rays in
the Galaxy this implies a short range propagation scale ($\sim 1$ kpc)
for high 
energy electrons, hence local sources are expected to play an
important role  (e.g. \cite{kob04,aha95,del10}).
Furthermore, a small number of sources well localized in space may
induce features in the spectral shape of the
electron energy spectrum \cite{del10,nis80} and anisotropies in the
arrival direction of very high energy electrons
(e.g.~\cite{kob04}). However, Delahaye et al.~\cite{del10} noted that a
full relativistic treatment of the energy losses smooths
the global spectral shape significantly reducing the peaking
structures resulting from using the Thomson approximation of the
inverse Compton energy losses.

\subsection{The ``all-electron'' spectrum}

Detection of high-energy energy electrons has been conducted over the
years employing the particle cascades produced by electrons in
calorimeters. Such approach provides sufficient energy resolution and
acceptance to extend the measurement of the
electron spectrum beyond 1~TeV. However, negative particles cannot be
straightforwardly separated  
from positive ones, hence the energy spectrum refers to the sum of 
\el + \ps (i.e. ``all-electron'').

\begin{figure}[ht]
\includegraphics[width=25pc]{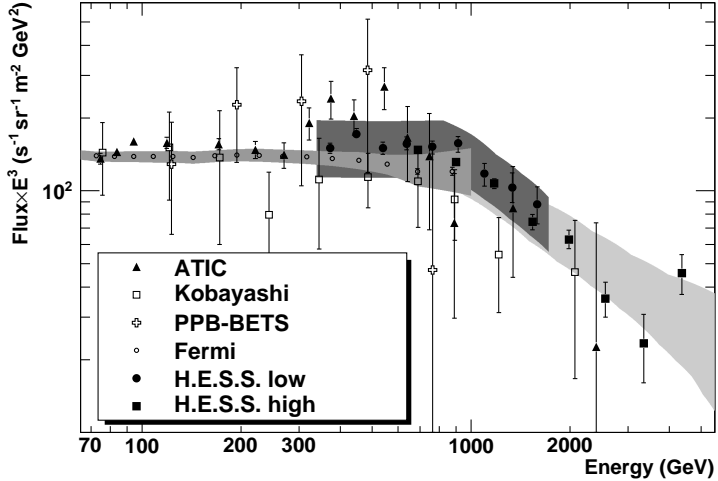}\hspace{2pc}
\caption{The all-electron energy spectrum measured by 
Kobayashi\protect\cite{kob99},
PPB-BETS~\protect\cite{tor08},
ATIC~\protect\cite{cha08}, 
H.E.S.S.~\protect\cite{aha08,aha09} and Fermi~\protect\cite{ack10}. The shaded
areas indicate the systematic uncertainties in the Fermi and H.E.S.S.
results (two sets of data measurements for H.E.S.S.).
\label{fig:fluxel2}}   
\end{figure}

As previously noted, recently the balloon-borne experiment ATIC
reported an excess of 
galactic cosmic-ray electrons at energies between 300-800
GeV~\cite{cha08}. The excess appeared as a structure in the
all-electron energy spectrum with a sharp cutoff around 620 GeV, which
was interpreted in terms of contribution of a nearby
astrophysical object such as pulsar or dark matter signal. The same
energy region was originally probed by the balloon-borne experiment by
Kobayashi et al.~\cite{kob99} and later on by the balloon-borne
experiment PPB-BETS~\cite{tor08}, by the satellite-borne experiment
Fermi~\cite{ack10,abd09} and by the atmospheric Cherenkov experiment
H.E.S.S.~\cite{aha09,aha08}.  
Figure~\ref{fig:fluxel2}\footnote{As commonly done, the figures in this
  paper show the 
  fluxes multiplied by E$^{3}$, where E is the energy in GeV. Reducing
  the decades of variation of the flux, this
  allows for a clearer picture of the spectral shapes. However,
  this implies that the absolute energy uncertainties are added to the
  flux uncertainties.} shows
these measurements. 
It can be noted that the recent measurements by Fermi and H.E.S.S. are
consistent with ATIC results within statistical and systematic
uncertainties but neither confirms this structure in the spectrum.  

All these measurements required powerful particle identification to
separate electrons from a vast background of protons and heavier
nuclei. The identification was based upon calorimetry information (in
the H.E.S.S. case the calorimeter was the atmosphere itself) such as
energy losses and shower development. Thickness of the employed
calorimeter as well as validity of simulations 
are fundamental ingredients for a reliable hadron rejection and
corresponding systematic uncertainties. Moreover, atmospheric
secondaries, i.e.  
produced by interaction of cosmic rays with the residual 
atmosphere above the payload, comprise an irreducibly
contamination for balloon-borne experiments. These secondary electrons
(and positrons) can only be estimated by Montecarlo or analytical 
calculations. Furthermore, since they approximately maintain the
spectral shape of the parent cosmic rays, their contamination
increases as the energy increases amounting often to 10\% or more of 
the signal in the hundred GeV region (e.g. see~\cite{tor08}).

Additional sources of discrepancies 
arise from 
efficiency and energy determinations. 
Selection efficiencies are an experimental challenge since they
require a very good knowledge of the detector performances during
data taking. Often, to reduce the systematic uncertainties, the
selection efficiencies are derived from flight data, but a fully
unbiased cross calibration of the efficiencies in flight is quite
impossible and simulations have to be used. The simulations are
usually validated by comparisons with test-beam data, which do not
account for the flight condition, and, whenever possible, flight
data. However, unproven assumption have to be made resulting in 
uncertainties that have to be included in the results.
It has to be noted that efficiency uncertainty usually affects the
absolute normalization of the fluxes and have a smaller impact on the
shape of the spectra. 

The shaded areas in Fig.~\ref{fig:fluxel2} shows the systematic
uncertainties in the Fermi and H.E.S.S. measurements. In the case of
Fermi they account for uncertainties in the estimation of the hadronic
background. However, these
uncertainties do not account for those deriving from the energy
determination. For the experiments in   
Fig.~\ref{fig:fluxel2}, the energy was obtained by measuring the
development of the electromagnetic cascade in the calorimeter. Also in
this case the thickness of the calorimeter plays a significant role in
the precision of the measurement with an energy resolution ranging
from $\sim 15\%$ for Fermi to $\sim 3\%$ for ATIC at few hundred
GeV. 

In the case of H.E.S.S. the shaded areas indicate the approximate systematic uncertainties arising from the modeling of hadronic interactions and of the atmosphere. The energy scale can approximately shift of 15\%. 
It should be mentioned that the energy determination strongly depends on simulations since experimental calibration at beam test cannot be performed, differently from space and balloon-borne experiments. 
Additionally, a contamination by the diffuse
$\gamma$-ray background affects the H.E.S.S. 
results,
estimated at a level of 10\% even if a significantly larger
contamination of $\simeq 50\%$ cannot be excluded~\cite{aha08}. It has
to be noted that the presence of this contamination implies that the
H.E.S.S. electron spectrum is probably an upper limit of the real spectrum and
that the contamination, due to the different spectral shapes of
electrons and diffuse $\gamma$-rays, is energy
dependent, hence potentially 
affecting the shape of the spectrum.

\begin{figure}[ht]
\includegraphics[width=25pc]{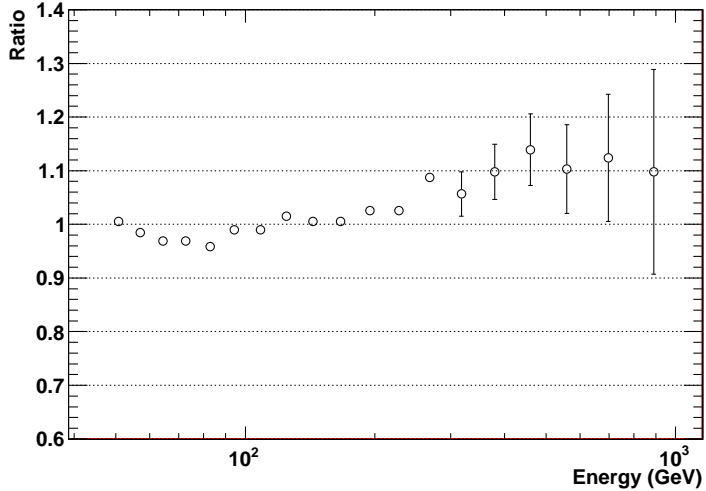}\hspace{2pc}
\caption{The ratio between the all-electron spectra measured by
  Fermi~\cite{ack10} using two event selections with different energy resolution (fluxes with better resolution divided by the ones with lower resolution).
\label{fig:Fermi}}   
\end{figure}

The effect of systematic uncertainties can be understood also looking
at Fig.~\ref{fig:Fermi} that shows the 
ratio between the all-electron spectra measured by
  Fermi~\cite{ack10} using two different event selections: one is the
  standard analysis published in~\cite{abd09}, the other uses a
  significantly smaller set of events ($\sim 5\%$ of the total) that
  crossed at least 12 radiation lengths in the calorimeter
  (see~\cite{ack10} for more details). The energy resolution of this
  second set of events is significantly better: about 3\%-5\% between
  100~GeV and 1~TeV. 
Considering that, on top of the statistical errors, the systematic
uncertainties are $\sim 5\%$, increasing as the energy increases and
without accounting for the absolute energy uncertainty
(see
Fig.~\ref{fig:fluxel2}), the two results are perfectly
consistent. However, it can also be noticed that the spectrum obtained
with the higher energy resolution is systematically higher than the one
with larger statistics above 200~GeV. This is another 
confirmation that a detailed interpretation of the electron data
require a careful analysis of the experimental data including all
statistical and systematic uncertainties. It also points to the need
for more precise measurements. 

As can be seen from Fig.~\ref{fig:fluxel2}, the ``all electron''
spectrum is well represented by a single power law of about -3 of
spectral index up to $\sim 1$ TeV, above which energy a cutoff like
feature is present.
A combination of contributions from a limited number of nearby sources
and distant ones uniformly distributed appears consistent with the
experimental results as shown in Fig. 14 of~\cite{del10}. 
However, as discussed, taking into account all experimental 
uncertainties spectral features in the hundred GeV region cannot
be excluded. 

\subsection{The \el spectrum}

\begin{figure}[ht]
\includegraphics[width=7.5cm,height=7.5cm]{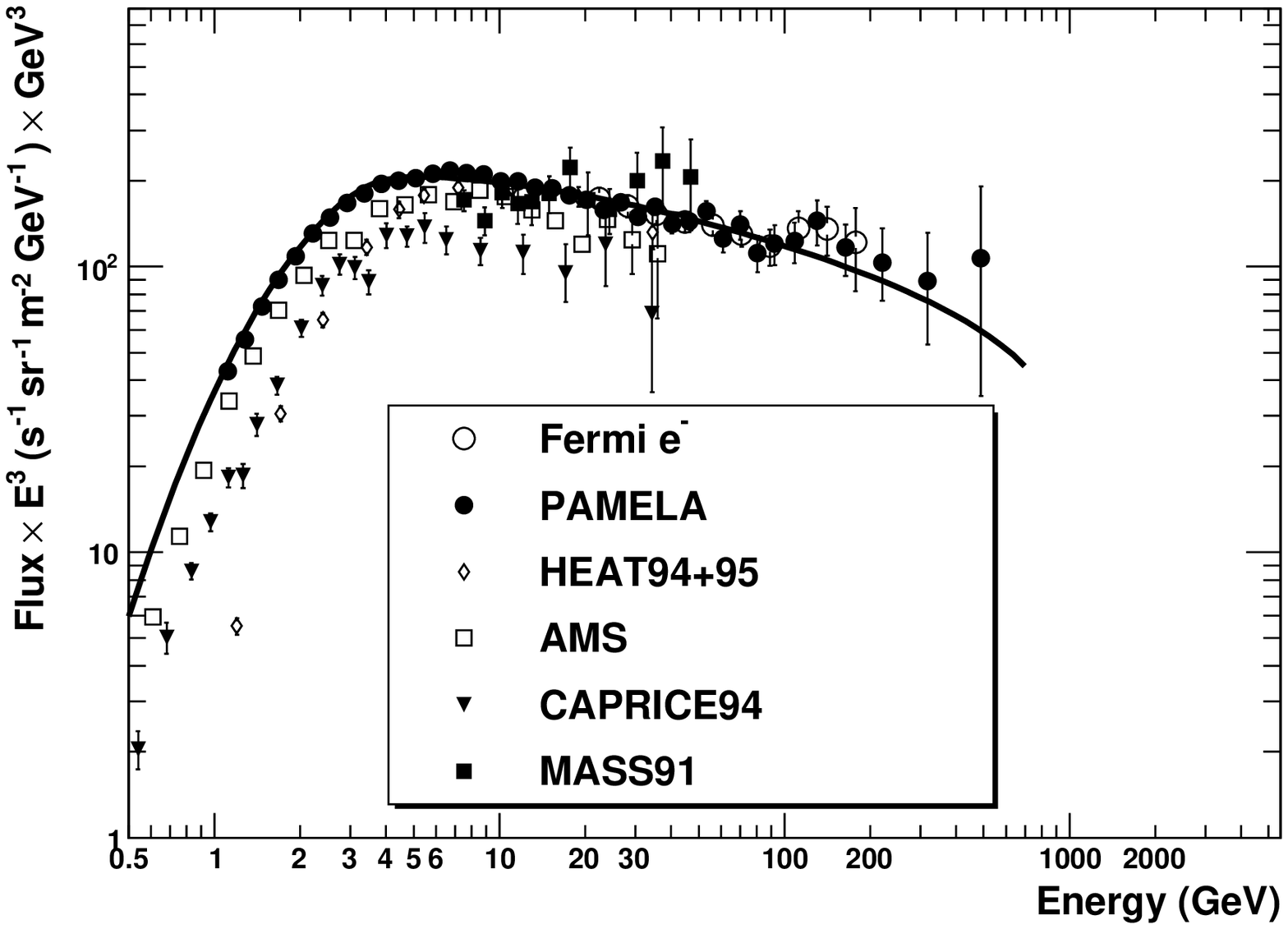}
\hspace{0.5cm}
\includegraphics[width=7.5cm,height=7.5cm]{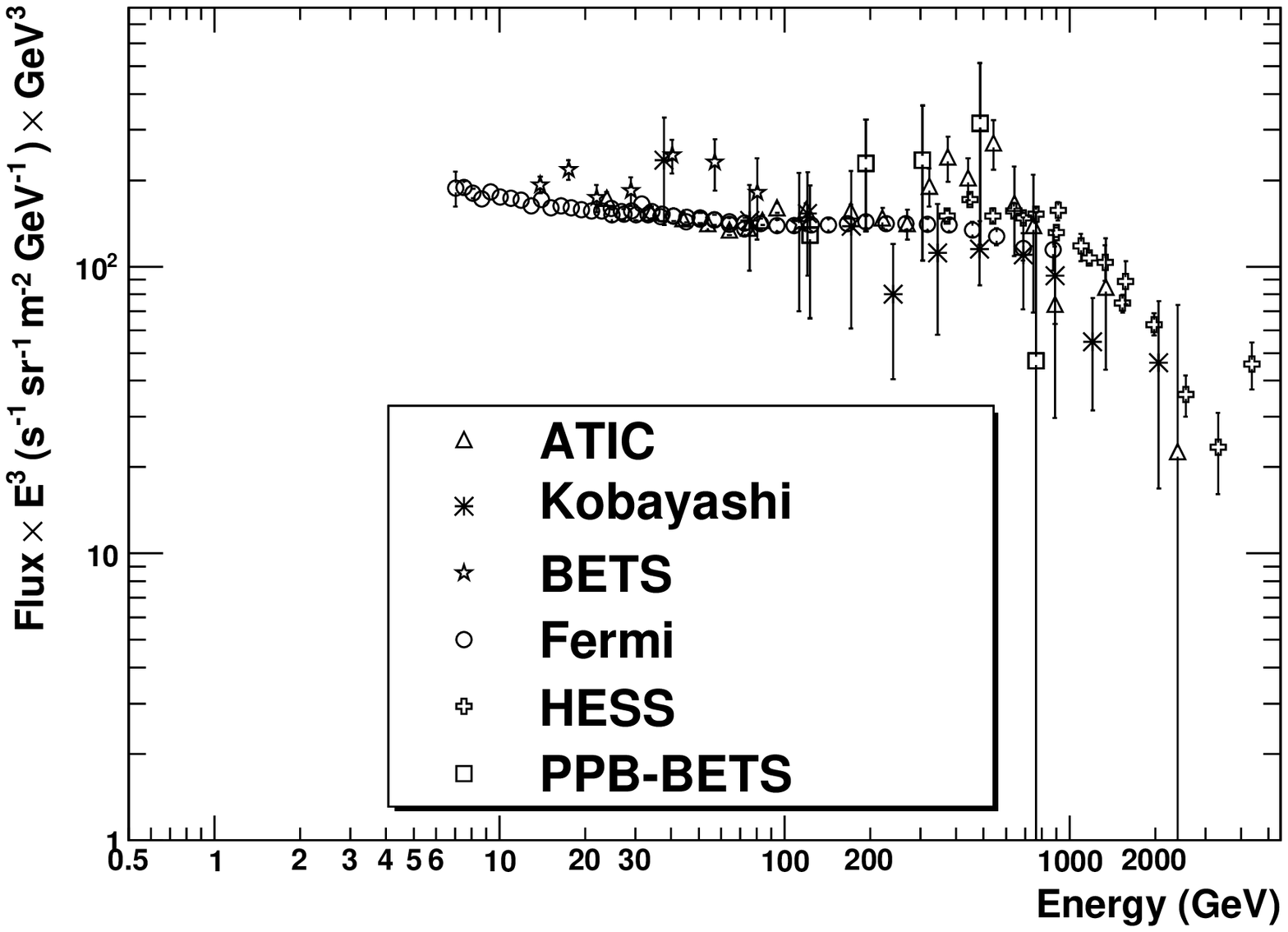} 
\caption{Left: recent measurements of the cosmic-ray \el energy spectrum:
CAPRICE94~\protect\cite{boe00}, 
HEAT~\protect\cite{duv01},
AMS~\protect\cite{alc00b}, MASS91~\protect\cite{gri02},
PAMELA~\protect\cite{adr11b} and Fermi~\protect\cite{ack12}.
The solid line is  a 
theoretical calculation based on the GALPROP code~\protect\cite{str98}. 
Right: the all-electron energy spectrum measured by 
Kobayashi\protect\cite{kob99}, BETS~\protect\cite{tor01}
PPB-BETS~\protect\cite{tor08},
ATIC~\protect\cite{cha08}, 
H.E.S.S.~\protect\cite{aha08,aha09} and Fermi~\protect\cite{ack10}.
\label{fig:fluxel1}}   
\end{figure}

It has been often noted, e.g.~\cite{del10}, that the separate
cosmic-ray \el and
\ps fluxes yield much more information and provide stronger constrains
to theoretical models than the all-electron
spectrum.
Since the first detection of cosmic-ray \el in the early
sixties~\cite{ear61}, several experiments were performed to measure
this component.
Figure~\ref{fig:fluxel1} left shows the \el energy spectrum measured 
by recent cosmic-ray experiments~\cite{adr11b,boe00,alc00b,duv01,gri02}
(the highest data point from HEAT~\cite{duv01} refer to the sum of electrons and positrons). 
and a 
theoretical calculation of the \el spectrum based on the GALPROP
code~\cite{str98}. 
The calculation (solid line) was performed using a spatial Kolmogorov
diffusion with spectral  
index $\delta = 0.34$ and diffusive reacceleration characterized 
by an Alfven speed $v_{A} = 36$~km/s, the halo height
was 4 kpc (parameters
from~\cite{ptu06}). The resulting flux 
was normalized to PAMELA data at $\sim 70$~GeV. For the
secondary \el production during propagation it used primary proton
and helium spectra reproducing the corresponding measured PAMELA
spectra~\cite{adr11a} and it was 
calculated for solar minimum,
using the force field approximation \cite{gle68} ($\Phi = 600$~MV).
As a comparison the Figure~\ref{fig:fluxel1} right shows 
measurements of the all-electron spectrum 
\cite{tor08,cha08,kob99,ack10,aha09,aha08,tor01} with the same energy scale. 

Differences in the data at low energies are mostly due to the effect
of solar modulation for the various data taking periods.   
Therefore, here we will mostly discuss measurements of 
electrons with energies above 10
GeV, for which these effects are much
less significant.

Especially the PAMELA and Fermi results  
show a rather smooth energy dependence of
the energy spectra in a relatively good agreement with the GALPROP
calculation except at higher energies where all experimental spectra
are harder. Such hardening can be explained with additional leptonic
components with a hard spectrum (e.g.~\cite{del10,ack10}), which, 
contributing 
equally to electrons and positrons, would likewise explain the 
the increase in the positron fraction measured by
PAMELA~\cite{adr09b}. The spectral flattening can be as
well reproduced by 
standard cosmic-ray models 
by adjusting the injection spectrum at the source;
however, these models cannot explain the PAMELA positron data. A
possible solution was proposed by Blasi~\cite{bla09b}, 
Ahlers et al.~\cite{ahl09} and Fujita et al.~\cite{fuj09} that 
considered the 
production and acceleration of secondaries electrons and positrons by
hadronic interactions of the accelerated protons in SNR shock waves.
With such assumption, they were able to fit the Fermi (and H.E.S.S.)
electron data and, at the same time, reproduce 
the PAMELA positron fraction. 
It should be mentioned that GALPROP does not fully describe cosmic-ray
electron propagation. This calculation is commonly used
assuming a continuous distribution of sources in the Galaxy.
However this
does not seem plausible for primary high energy electrons, 
since, as previously discussed,  this
assumption should only hold for a relatively close
neighborhood. Furthermore, as pointed out in \cite{sha09}, 
there is a higher concentration of SNRs 
in the spiral arms of the Galaxy, therefore one should consider an
inhomogeneous source distribution. 

\begin{figure}[ht]
\includegraphics[width=25pc]{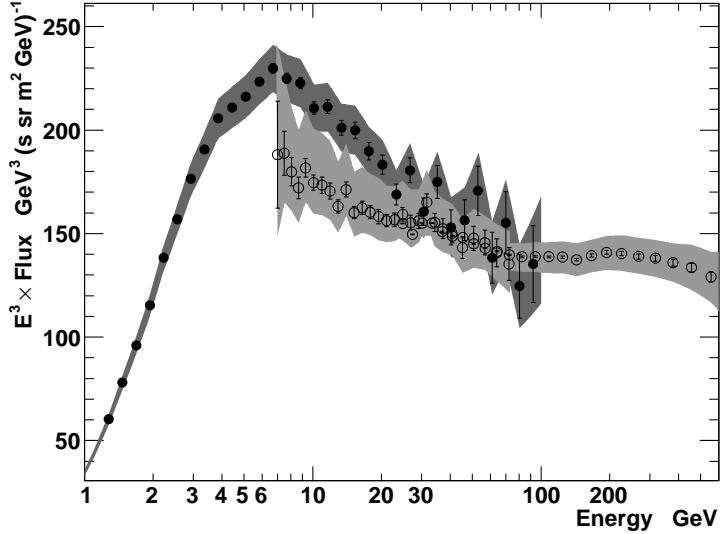}\hspace{2pc}
\caption{The all (\el + \ps) electron energy spectrum from
  Fermi~\protect\cite{ack10} (open circles) and 
  PAMELA~\protect\cite{adr11b,adr10a} (full circles).  
\label{fig:fluxFePam}}   
\end{figure}

As in the case of the very-high-energy all-electron spectrum the
interpretation of the results depends on the precision of the
measurements. Especially systematic uncertainties, often larger than
statistical errors, have to be properly taken into account in any data
interpretation. However, it is very difficult to correctly estimate
systematic uncertainties. An excellent approach to their determination
resides in comparing measurements taken with different apparatus.
For example, we compared PAMELA electron spectrum to the Fermi
all-electron spectrum. For this comparison, 
we constructed a PAMELA all-electron spectrum
using the PAMELA \el data~\cite{adr11b} and positron
fraction~\cite{adr10a}.
 Figure~\ref{fig:fluxFePam} 
shows the Fermi and PAMELA all-electron spectra up to 100~GeV, i.e. the
highest energy bin for which the PAMELA collaboration presented the
positron fraction. The PAMELA spectrum appears 
softer than the Fermi one, however the two spectral indexes differ by
less than a standard deviation ($-3.17 \pm 0.07$ for PAMELA and
$-3.112 \pm 0.002$ for Fermi) when fitting the data with a single
power-law from 30 to 100 GeV and accounting only for statical
errors. 
At lower energies, the electron flux measured by PAMELA is higher
(about 20\% at 10 GeV) than that measured by Fermi. Considering that
the data were collected partially over the same period of time the
differences cannot be ascribed to solar modulation. 
However, 
systematic uncertainties, indicated as shaded areas in
Fig.~\ref{fig:fluxFePam}, account for most of the
differences. Once more this illustrates why systematic
uncertainties cannot be neglected when modelling experimental
results. 
 
\begin{figure}[ht]
\includegraphics[width=25pc]{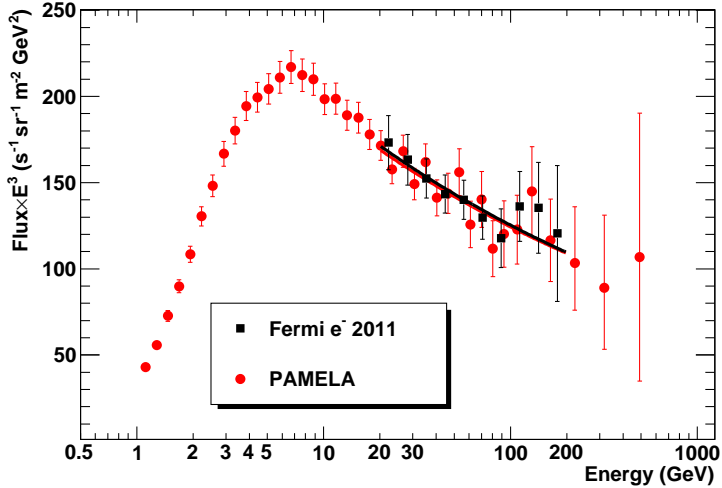}\hspace{2pc}
\caption{The \el  energy spectrum measured by
  Fermi~\protect\cite{ack12} and 
  PAMELA~\protect\cite{adr11b}. The solid lines are single power-law
  fit to the data between 20 and 200 GeV (red: PAMELA, black: Fermi). 
For both measurements the error bars 
  include both statistical 
  and systematic uncertainties quadratically summed and 
these were the errors considered by the
fitting procedures.
\label{fig:fluxFePam2}}   
\end{figure}

Recently the Fermi collaboration published data on \el and \ps fluxes
\cite{ack12} 
where the two species were distinguished up to 200 GeV using the
opposite distortion 
of the Earth's shadow cased by Earth's magnetic
field. These results are particularly interesting not only because
they confirm the PAMELA positron results but also because the
differences between the Fermi and PAMELA \el spectra are significantly
reduced respect to the comparison discussed above 
(Fig.~\ref{fig:fluxFePam}).  
Figure~\ref{fig:fluxFePam2} 
shows the \el spectrum measured by Fermi compared with the one 
by PAMELA~\cite{adr11b} along with single power-law
  fits to the data between 20 and 200 GeV.  
An excellent
  agreement can be noticed: there are negligible differences in the absolute
values and in 
the two spectral index  
($-3.19 \pm 0.04$ for PAMELA and
$-3.20 \pm 0.08$ for Fermi). Apparently, the better agreement
between PAMELA and Fermi electron spectra is due to improved
instrument 
response functions used in the Fermi data (see~\cite{ack12}). This
agreement can be used to reduce the effect of systematic
uncertainties, providing an \el spectrum more precise than each single
measurement. It is interesting to note that both PAMELA
and Fermi \el spectra harden and hint to a structure around 100~GeV
not unlike a similar feature noticeable in the ATIC all-electron
spectrum (see Fig.~\ref{fig:fluxel2}). 
 
\subsection{Future}
Experiments based on atmospheric Cherenkov telescopes such as
H.E.S.S. and the future CTA~\cite{cta11} are very important for
studies of the 
Galactic cosmic-ray electron spectrum since they are able to probe the
TeV energy region with orders of magnitude larger collection areas
than balloon- and satellite-borne experiments. However, the
uncertainties related to hadronic and diffuse $\gamma$-ray
backgrounds and to the energy determinations significantly overshadow the
statistical uncertainties muddling the interpretation of the
measurements. The authors believe that a direct measurement of the
cosmic-ray electron spectrum is still the preferable approach also in
the high-energy region. Future experiments such as the space-borne
CALET and Gamma-400 and balloon-borne CREST may provide the needed
precision for a significantly improved understanding of the cosmic-ray
electron spectrum, their propagation and origin. 

CALET~\cite{tor11} is an experiment designed to measured the
all-electron cosmic ray spectrum 
from 1 GeV to 20 TeV. The apparatus is built around a 30 radiation length
calorimeter and it 
will be placed on board the International Space Station (ISS) sometime
around 2014. With an 
acceptance of about 0.12 m$^{2}$sr, CALET will significantly extend
previous electron measurement with a significant improvement on the
systematic uncertainties.

A similar
calorimetry approach will be employed by the Gamma-400~\cite{gal12}
experiment. This experiment is aimed 
to study the
high-energy gamma-ray flux and cosmic-ray  electrons and nuclei.
The apparatus will be placed on board a Russian satellite, which launch
is foreseen for 2017-2018. With a similarly deep but significantly
larger calorimeter (acceptance of about 1 m$^{2}$sr), Gamma-400
should 
be able to extend the cosmic-ray measurements performed
by CALET.

A different approach has been proposed for the CREST
experiment~\cite{sch08} that
aims to detect the 
synchrotron photons generated at x-ray energies by TeV cosmic-ray
electrons in the Earth's magnetic field. While affected by a
relatively poor energy resolution, the experiment, sensitive to
electrons of energies greater than 2~TeV, 
can efficiently 
sample the multi-TeV region.
The apparatus, carried on a long duration
balloon-flight\footnote{An engineering flight of CREST took place in
  2009~\cite{nut09}, a long duration balloon flight has not yet been
  scheduled.}, will be able to observe up to 30 electrons with energy 
greater 
than 2 TeV in a 2 week flight~\cite{sch08}. 

As previously pointed out, separate data of positrons and electrons
provide stronger grounds to any interpretation of the cosmic-ray
electron data. Till now the \el spectrum has been measured up to
600~GeV 
thanks to the PAMELA data~\cite{adr11b} but with limited
statistics. However, more precise data can be expected in the future. 
In fact, on the 19th of May 2011 the AMS-02 apparatus~\cite{bat05} was
installed onboard the 
ISS and it started collecting data. The
apparatus is  equipped with a permanent
magnet, a silicon tracking device, an electromagnetic
calorimeter, a Transition Radiation
Detector and a Ring Imaging Cherenkov detector. Similar in scope to
the PAMELA experiment, AMS 
has a significantly larger acceptance (about
a factor 20) and additional particle identifier 
detectors such as a Transition Radiation
that will provide
a significant improvement in statistics and systematics respect to
PAMELA. Considering the 10-year long AMS mission, the experiment may
be able to provide sufficiently
precise results to search for structures in the electron
spectrum at least up to 1~TeV.